\documentclass[aps,prstab,floatfix,showpacs,amsmath,amssymb]{revtex4-2}
\usepackage{graphicx}
\usepackage{dcolumn}
\usepackage{bm}
\begin{document}

\title{Effect of solenoid lens field on electron beam emittance}
\author{Kazuaki Togawa}
\email{togawa@spring8.or.jp}
\affiliation{RIKEN SPring-8 Center, 1-1-1 Kouto, Sayo, Hyogo 679-5148, Japan}
\date{\today}

\begin{abstract}

In an injector system of an X-ray free electron laser (XFEL), solenoid lenses are typically used to confine low-emittance electron beams to low-energy region below a few MeV. Because non-thermionic emittance at such a low-energy region is easily deteriorated by nonlinear electromagnetic fields, it is important to determine the properties of a solenoid lens on electron beam emittance in the design of XFEL injectors. We derived an approximate solution to emittance growth due to lens aberration by a paraxial approximation. It was found that the derivative of the longitudinal magnetic field strongly affects beam emittance, and its growth is proportional to the fourth power of the beam radius. Various properties of the beam can be analyzed as long as the longitudinal magnetic field distribution is prepared using a simulation or measurement. In this study, a theoretical procedure to obtain the emittance growth in the solenoid lens is introduced and the design considerations of the solenoid lens of the SACLA injector are described.

\end{abstract}

\pacs{07.55.Db, 41.60.Cr}

\maketitle

\section{Introduction}

X-ray free electron lasers (XFELs) require an extremely fine electron beam whose emittance is below 1 $\mu$m to lase high-intensity coherent X-ray beams \cite{McNeil01}. Two types of electron guns are used currently for XFELs: photo-cathode radio-frequency and thermionic-cathode high-voltage guns \cite{PhysRevSTAB.11.030703, PhysRevSTAB.10.020703}. Because the electron beams must be accelerated and transported in the low-energy region from 0 to a few MeV in either case, the non-thermionic emittances are easily deteriorated by nonlinear electromagnetic fields such as space charge and accelerating fields.

At a low-energy region, a series of solenoid coils or lenses are frequently used to confine the electron beam because of the strong space charge force that results in the beam spreading out instantly. It is common that the solenoid lens has an intrinsic aberration property. This property can determine the limit of imaging resolution in electron microscopes \cite{RevModPhys.59.639}. In electron accelerators, it causes emittance growth that deteriorates XFEL output power. Therefore, it is important to evaluate the basic properties of a solenoid lens to minimize emittance growth in electron injector design. In this study, an approximate solution of emittance growth in the solenoid lens is derived and the design consideration for the low-emittance injector of the Japanese XFEL facility, SACLA \cite{Ishikawa01}, is described.

\section{Emittance growth in solenoid lens}

\subsection{Equation of motion in axisymmetric magnetic field}

The equation of motion of an electron in an axisymmetric magnetic field with cylindrical coordinates is given by
	\begin{eqnarray}
	\gamma m_e\frac{dv_r}{dt}&=&\frac{\gamma m_ev_\theta^2}{r}-ev_\theta B_z, \label{eqn:motion1} \\
	\gamma m_e\frac{d(rv_\theta)}{dt}&=&-er(v_zB_r-v_rB_z), \label{eqn:motion2}
	\end{eqnarray} 
where $\gamma$ is the relativistic mass factor, $m_e$ is the electron mass, $e$ is the elementary charge, $v_\theta$ and $v_z$ are the rotational and longitudinal components of velocity, respectively, and $B_r$ and $B_z$ are the radial and longitudinal components of the magnetic field, respectively. Here, the rotational component $B_\theta$ that is induced by the beam current is ignored and $\gamma$ is assumed to be constant. 

The velocity increase in the radial direction is obtained by integrating Eq.(\ref{eqn:motion1}):
	\begin{eqnarray}
	\Delta v_r=\int\biggl(\frac{v_\theta^2}{r}-\frac{e}{\gamma m_e}v_\theta B_z\biggr)dt. \label{eqn:dvr1}
	\end{eqnarray} 
Using $dz=v_zdt$ and $v_z\approx const.$, Eq.(\ref{eqn:dvr1}) can be rewritten as
	\begin{eqnarray}
	\Delta v_r=\frac{1}{v_z}\int\biggl(\frac{v_\theta^2}{r}-\frac{e}{\gamma m_e}v_\theta B_z\biggr)dz. \label{eqn:dvr2}
	\end{eqnarray} 
Next, we derive the relation between the velocity in the rotational direction $v_\theta$ and rotational component of vector potential $A_\theta$. From $\overrightarrow{\rm B}=rot\overrightarrow{\rm A}$, $B_r$ and $B_z$ can be expressed as
	\begin{eqnarray}
	B_r&=&-\frac{\partial A_\theta}{\partial z}, \label{eqn:Br} \\
	B_z&=&\frac{1}{r}\frac{\partial(rA_\theta)}{\partial r}, \label{eqn:Bz}
	\end{eqnarray} 
where $\partial A_z/\partial \theta=\partial A_r/\partial \theta=0$ is used because the magnetic field is axisymmetrical. Using Eq.(\ref{eqn:Br}) and (\ref{eqn:Bz}), Eq.(\ref{eqn:motion2}) becomes
	\begin{eqnarray}
	\gamma m_e\frac{d(rv_\theta)}{dt}=e\frac{d(rA_\theta)}{dt}. \label{eqn:motion2b}	
	\end{eqnarray} 
When $A_\theta$ is zero at the initial point, i.e., the magnetic field does not penetrate a beam emitter cathode, $v_\theta$ can be expressed as
	\begin{eqnarray}
	v_\theta=\frac{e}{\gamma m_e}A_\theta. \nonumber
	\end{eqnarray} 
Substituting $v_\theta$ in Eq.(\ref{eqn:dvr2}), $\Delta v_r$ becomes
	\begin{eqnarray}
	\Delta v_r=\frac{1}{v_z}\biggl(\frac{e}{\gamma m_e}\biggr)^2\int\biggl(\frac{A_\theta^2}{r}-A_\theta B_z\biggr)dz. \nonumber
	\end{eqnarray} 
As a result, the variation of the electron diverging angle $\Delta r'$ after passing through the solenoid field becomes:
	\begin{eqnarray}
	\Delta r'=\frac{\Delta v_r}{v_z}=\biggl(\frac{e}{\beta\gamma m_ec}\biggr)^2\int\biggl(\frac{A_\theta^2}{r}-A_\theta B_z\biggr)dz, \label{eqn:drd}
	\end{eqnarray} 
where $\beta=v_z/c$.

\subsection{Paraxial expansion of magnetic field}

To obtain the paraxial approximation of $A_\theta$ and $B_z$, we start with the Maxwell equation:
	\begin{eqnarray}
	\nabla^2\overrightarrow{\rm A}=0. \nonumber
	\end{eqnarray} 
The rotational component can be written explicitly as
	\begin{eqnarray}
	\frac{1}{r}\frac{\partial}{\partial r}\biggl(r\frac{\partial A_\theta}{\partial r}\biggr)+\frac{\partial^2A_\theta}{\partial z^2}-\frac{A_\theta}{r^2}=0, \label{eqn:mxwl2}
	\end{eqnarray} 
where $\partial A_\theta/\partial\theta=0$ is used because the magnetic field is axisymmetrical. Moreover, $A_\theta$ can be expanded as
	\begin{eqnarray}
	A_\theta(r,z)=\sum^{\infty}_{n=0}a_n(z)r^n. \label{eqn:atheta}
	\end{eqnarray} 
Substituting Eq.(\ref{eqn:atheta}) in Eq.(\ref{eqn:mxwl2}), we obtain the relation of the coefficients $a_n$: 
	\begin{eqnarray}
	\sum^{\infty}_{n=0}(1-n^2)a_nr^{n-2}=\sum^{\infty}_{n=0}a_n^{(2)}r^n, \label{eqn:poly1}
	\end{eqnarray} 
where $a_n^{(k)}$ shows $k$th derivative of the coefficient:
	\begin{eqnarray}
	a_n^{(k)}=\frac{d^ka_n}{dz^k}. \nonumber
	\end{eqnarray} 
By comparing each coefficients of $r^n$ in Eq.(\ref{eqn:poly1}), we obtain
	\begin{eqnarray}
	a_0=0, \nonumber
	\end{eqnarray} 
	\begin{eqnarray}
	[1-(n+2)^2]a_{n+2}=a_n^{(2)}. \label{eqn:poly2}
	\end{eqnarray} 
Eq.(\ref{eqn:poly2}) can be rewritten as
	\begin{eqnarray}
	a_{n+2}=-\frac{a_n^{(2)}}{(n+1)(n+3)}. \nonumber
	\end{eqnarray} 
From $a_0=0$, all of the coefficients with even index numbers become zero. The coefficients with odd index numbers are
	\begin{eqnarray}
	a_1&=&a_1, \nonumber \\
	a_3&=&-\frac{a_1^{(2)}}{2\cdot4}, \nonumber \\
	a_5&=&-\frac{a_3^{(2)}}{4\cdot6}=(-1)^2\frac{1}{2\cdot4\cdot4\cdot6}a_1^{(4)}, \nonumber \\
	a_7&=&-\frac{a_5^{(2)}}{6\cdot8}=(-1)^3\frac{1}{2\cdot4\cdot4\cdot6\cdot6\cdot8}a_1^{(6)}, \nonumber \\
	&\cdot& \nonumber \\
	&\cdot& \nonumber \\
	&\cdot& \nonumber \\
	a_n&=&\frac{(-1)^{\frac{n-1}{2}}}{2^{n-1}\left[\left(\frac{n-1}{2}\right)!\right]^2\frac{n+1}{2}}a_1^{(n-1)}\ \ \ \ (n=3,5,7,\cdot\cdot\cdot). \nonumber
 	\end{eqnarray} 
Because $n$ is an odd number, it can be replaced by $2n+1$:
	\begin{eqnarray}
	a_{2n+1}=\frac{(-1)^n}{2^{2n}n!(n+1)!}a_1^{(2n)}\ \ \ \ (n=1,2,3,\cdot\cdot\cdot). \nonumber
 	\end{eqnarray} 
Consequently, the vector potential $A_\theta$ is expressed using the following polynomial:
	\begin{eqnarray}
	A_\theta(r,z)=\sum^{\infty}_{n=0}\frac{(-1)^n}{2^{2n}n!(n+1)!}a_1^{(2n)}(z)r^{2n+1}. \label{eqn:atheta2}
	\end{eqnarray} 
The longitudinal magnetic field $B_z$ can be expressed as a polynomial using Eq.(\ref{eqn:Bz}) and Eq.(\ref{eqn:atheta2}):
	\begin{eqnarray}
	B_z(r,z)=\sum^{\infty}_{n=0}\frac{(-1)^n}{2^{2n}(n!)^2}2a_1^{(2n)}(z)r^{2n}. \nonumber 
	\end{eqnarray}
Here, $a_1$ gives the longitudinal component of the z-axis magnetic field: 
	\begin{eqnarray}
	B_z(0,z)&=&B_{z0}(z)=2a_1(z). \nonumber 
	\end{eqnarray} 
Using $B_{z0}$, we obtain the final expressions of $A_\theta$ and $B_z$:
	\begin{eqnarray}
	A_\theta(r,z)&=&\sum^{\infty}_{n=0}\frac{(-1)^n}{n!(n+1)!}B_{z0}^{(2n)}(z)\left(\frac{r}{2}\right)^{2n+1}, \label{eqn:atheta3} \\
	B_z(r,z)&=&\sum^{\infty}_{n=0}\frac{(-1)^n}{(n!)^2}B_{z0}^{(2n)}(z)\left(\frac{r}{2}\right)^n. \label{eqn:bz3} 
	\end{eqnarray}

\subsection{Emittance growth owing to nonlinear focusing}

To evaluate beam focusing properties of the solenoid lens, $A_\theta$ and $B_z$ are approximated using the first and second terms in Eq.(\ref{eqn:atheta3}) and Eq.(\ref{eqn:bz3}), respectively: 
	\begin{eqnarray}
	A_\theta(r,z)&=&\frac{1}{2}rB_{z0}(z)-\frac{1}{16}r^3B''_{z0}(z), \label{eqn:atheta4} \\
	B_z(r,z)&=&B_{z0}(z)-\frac{1}{4}r^2B''_{z0}(z), \label{eqn:bz4}
	\end{eqnarray} 
where $B''_{z0}$ denotes the second derivative $B_{z0}^{(2)}$. Using Eq.(\ref{eqn:atheta4}) and Eq.(\ref{eqn:bz4}), the integrand in Eq.(\ref{eqn:drd}) becomes
	\begin{eqnarray}
	-\frac{r}{4}B_{z0}^2(z)+\frac{r^3}{8}B_{z0}(z)B_{z0}''(z)-\frac{3r^5}{256}B_{z0}''^2(z). \nonumber
	\end{eqnarray} 
The last term with the 5th order of $r$ is negligible. Substituting them in Eq.(\ref{eqn:drd}), the increase in the beam diverging angle $\Delta r'$ can be expressed as
	\begin{eqnarray}
	\Delta r'=\left(\frac{e}{\beta\gamma m_ec}\right)^2\left(-\int\frac{B_{z0}^2(z)}{4}dz\cdot r+\int\frac{B_{z0}(z)B_{z0}''(z)}{8}dz\cdot r^3\right). \label{eqn:drd2}
	\end{eqnarray} 
Using the following relation
	\begin{eqnarray}
	\int B_{z0}'^2(z)dz+\int B_{z0}(z)B_{z0}''(z)dz=[B_{z0}(z)B_{z0}'(z)]_{-\infty}^\infty=0, \nonumber
	\end{eqnarray} 
Eq.(\ref{eqn:drd2}) becomes
	\begin{eqnarray}
	\Delta r'=-\frac{1}{8}\left(\frac{e}{\beta\gamma m_ec}\right)^2\left(2\int B_{z0}^2(z)dz\cdot r+\int B_{z0}'^2(z)dz\cdot r^3\right). \label{eqn:drd3}
	\end{eqnarray} 
In an ordinary case, a field region created by the thin lens is short and the a radial position of electron $r$ does not change significantly. Therefore, we assume that $r$ is constant; however $r'$ changes in the solenoid field. Linear beam focusing is obtained from the first term, whereas the second term causes nonlinear beam focusing, i.e., lens aberration. It should be noted that the sign of the nonlinear term is always similar to that of the linear one. This means that the lens aberration cannot be compensated using another lens.

The focal length of the solenoid lens $f$ can be estimated using the first term of Eq.(\ref{eqn:drd3}):
	\begin{eqnarray}
	\frac{1}{f}=-\frac{\Delta r'}{r}=\frac{1}{4}\left(\frac{e}{\beta\gamma m_ec}\right)^2\int B_{z0}^2(z)dz. \label{eqn:flength} 
	\end{eqnarray} 
In a low energy case, using $(\beta\gamma)^2\approx2eV/(m_ec^2)$, the focal length is approximated as
	\begin{eqnarray}
	\frac{1}{f}=\frac{e}{8m_eV}\int B_{z0}^2(z)dz, \nonumber
	\end{eqnarray} 
where $V$ is the beam voltage. This is a familiar formula commonly described in electron optics textbooks \cite{Pierce01}.

Next, we consider the emittance growth owing to nonlinear focusing. The normalized rms emittance increase in the horizontal (or vertical) direction is given by
	\begin{eqnarray}
	\Delta\varepsilon_{n,rms}^x=\frac{\beta\gamma}{2}\sqrt{<r^2><\Delta r'^2>-<r\cdot\Delta r'>^2}, \label{eqn:emittance} 
	\end{eqnarray} 
where $< >$ denotes the mean values weighted by the current density. Each mean value can be calculated using Eq.(\ref{eqn:drd3}):
	\begin{eqnarray}
	<r^2>&=&\frac{\int J(r)r^2dS}{\int J(r)dS}=\frac{\int r^3 dr}{\int rdr}=\frac{r^2}{2}, \label{eqn:avr2} \\
	<\Delta r'^2>&=&\frac{\int J(r)\Delta r'^2dS}{\int J(r)dS}=\frac{\int(2k_1r+k_2r^3)^2rdr}{\int rdr}=2k_1^2r^2+\frac{4}{3}k_1k_2r^4+\frac{1}{4}k_2^2r^6, \label{eqn:avdr2} \\
	<r\cdot\Delta r'>&=&\frac{\int J(r)r\Delta r'dS}{\int J(r)dS}=\frac{\int(2k_1r+k_2r^3)r^2dr}{\int rdr}=k_1r^2+\frac{1}{3}k_2r^4, \label{eqn:avrdr} 
	\end{eqnarray}
where
	\begin{eqnarray}
	k_1&=&-\frac{1}{8}\left(\frac{e}{\beta\gamma m_ec}\right)^2\int B_{z0}^2(z)dz, \nonumber \\
	k_2&=&-\frac{1}{8}\left(\frac{e}{\beta\gamma m_ec}\right)^2\int B_{z0}'^2(z)dz, \nonumber
	\end{eqnarray}
and the beam current density is assumed to be uniform ($J(r)=const.$).
Substituting Eq.(\ref{eqn:avr2})-(\ref{eqn:avrdr}) in Eq.(\ref{eqn:emittance}), the emittance increase becomes
	\begin{eqnarray}
	\Delta\varepsilon_{n,rms}^{lens}=\frac{\beta\gamma}{2}\frac{|k_2|}{6\sqrt2}r^4. \nonumber
	\end{eqnarray} 
Finally, we obtain the approximative expression of the emittance growth in the solenoid lens:
	\begin{eqnarray}
	\Delta\varepsilon_{n,rms}^{lens}=\frac{1}{96\sqrt2\beta\gamma}\left(\frac{e}{m_ec}\right)^2\int\left(\frac{dB_{z0}(z)}{dz}\right)^2dz\cdot r^4. \label{eqn:emittance2}
	\end{eqnarray} 

This equation suggests the following two important features: 1) Emittance increases intensely when beam size becomes large. Beam size should be sufficiently small to maintain low emittance. In addition, the law that emittance growth is proportional to the fourth power of the beam radius appears in an electric field \cite{PhysRevAccelBeams.20.053401}. 2) The derivative of the longitudinal magnetic field affects beam emittance. Therefore, the magnetic field should vary gradually along the z-axis. This can be achieved by selecting a large aperture of the solenoid yoke.

\section{Solenoid lens design considerations for a low-emittance injector}

\begin{figure}[ht]
\includegraphics[width=15cm]{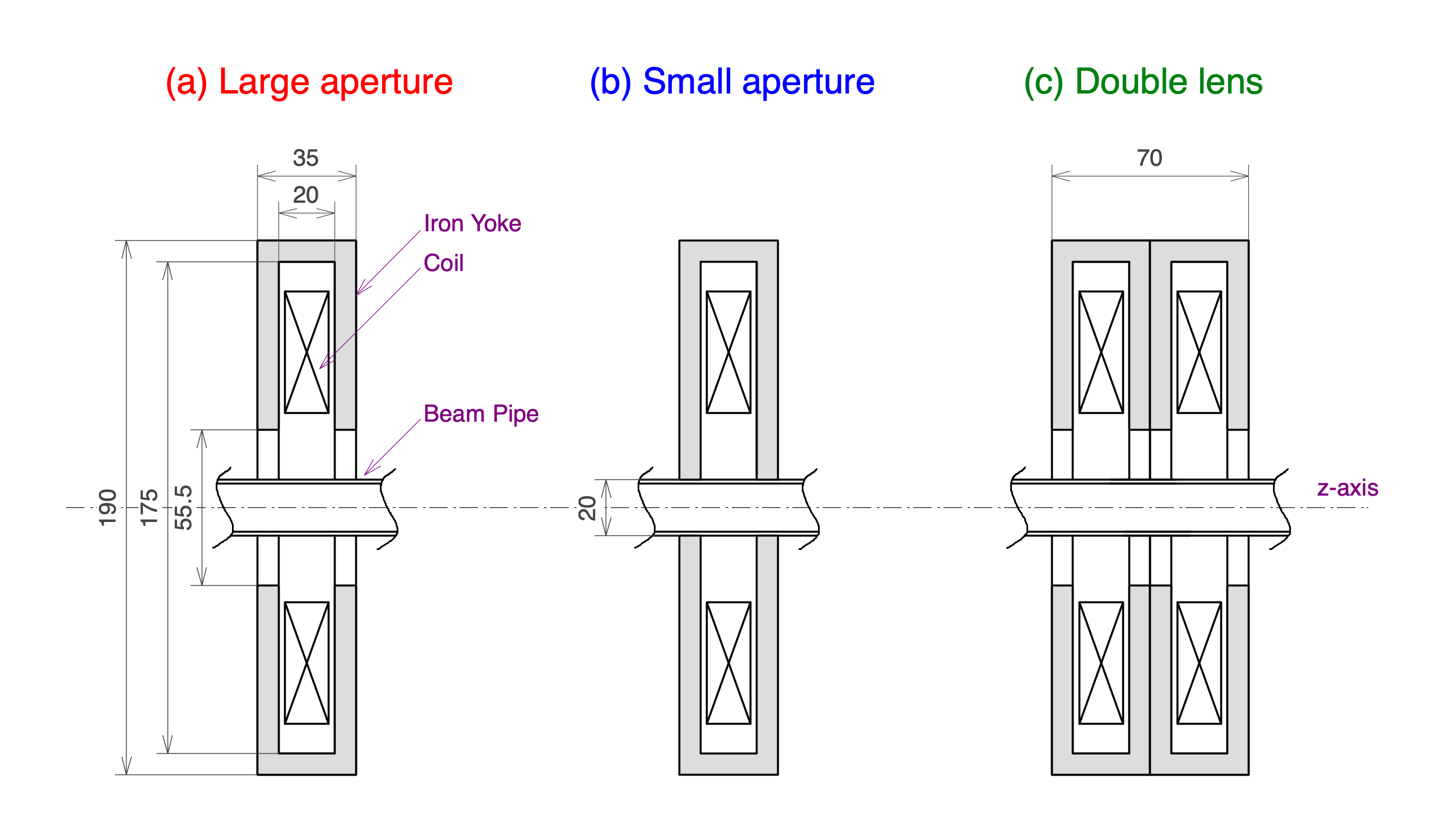}
\caption{\label{fig:shape} Cross section of the solenoid lenses with: (a) a large pole piece aperture of yoke, (b) a small pole piece aperture of yoke, and (c) a double lens whose magnetic field polarity are opposite. The dimensions are described in mm. The SCSS test accelerator and SACLA adopted type (a) with outer yoke diameters of 190 mm and 304 mm, respectively. }
\end{figure}

In the SACLA injector system and SCSS test accelerator \cite{Shintake03}, which was upgraded to SCSS+ \cite{Owada01}, a total of ten solenoid lenses are used to focus and transport the beam whose energy range is 0.5-1 MeV. Therefore, it is essential to minimize the non-thermionic emittance growth owing to the solenoid fields. To determine the shape of solenoid yoke, three types of solenoid lenses were analyzed. Type (a), which was chosen for the SCSS test accelerator and SACLA, has a large pole piece aperture of yoke. Type (b) has a small aperture, which was preferable in creating a strong magnetic field in a narrow space. Type (c) was composed of two solenoid lenses whose magnetic field polarity were opposite. The opposite fields can cancel the rotation around the beam axis. The cross section of the solenoid lenses are shown in Fig.~\ref{fig:shape}.

\begin{figure}[ht]
\includegraphics[width=10cm]{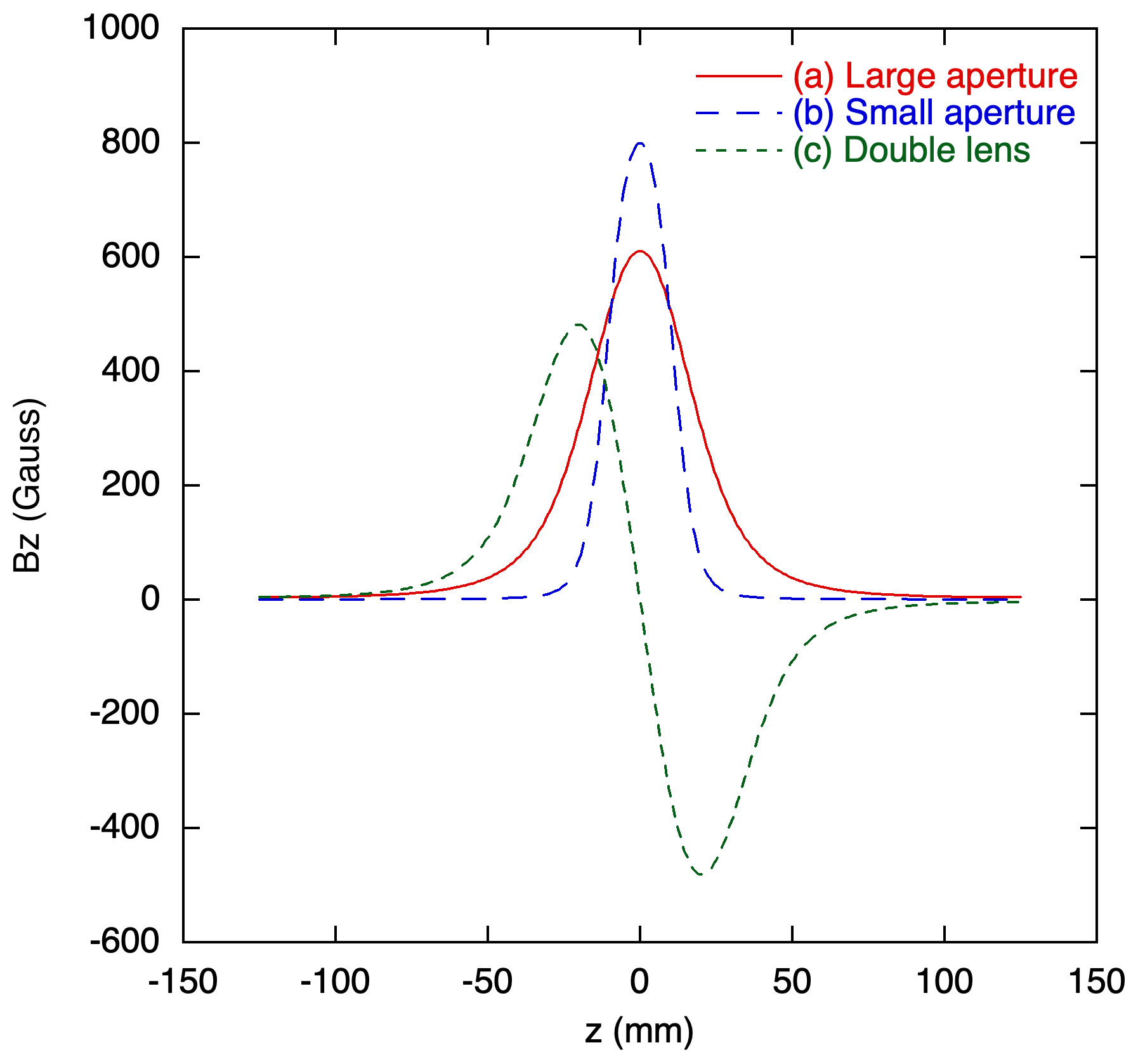}
\caption{\label{fig:field}Longitudinal magnetic field on z-axis calculated using the POISSON code.}
\end{figure}

Fig.~\ref{fig:field} shows the longitudinal magnetic field distribution on the z-axis calculated using the POISSON code \cite{POISSON}. The coil currents for each type were assumed to be 2297 A$\cdot$turn for (a), 1554 A$\cdot$turn for (b), and 4236 A$\cdot$turn for (c). A focul length of 0.3 m for a 500 keV beam was obtained from these currents, which was determined using Eq.(\ref{eqn:flength}). When the magnetic flux penetrates the cathode, normalized rms emittance increases by
	\begin{eqnarray}
	\Delta\varepsilon_{n,rms}^{mag}=\frac{eB_{zc}r_c^2}{8m_e c}, \nonumber
	\end{eqnarray}
where $B_{zc}$ is the longitudinal magnetic field on the cathode and $r_c$ is the cathode radius. The large pole piece aperture causes a wide magnetic field distribution. Therefore, the distance between the cathode and first solenoid lens must be long sufficiently to certify the small field. In SACLA, the distance was selected 135 mm. 

\begin{figure}[ht]
\includegraphics[width=10cm]{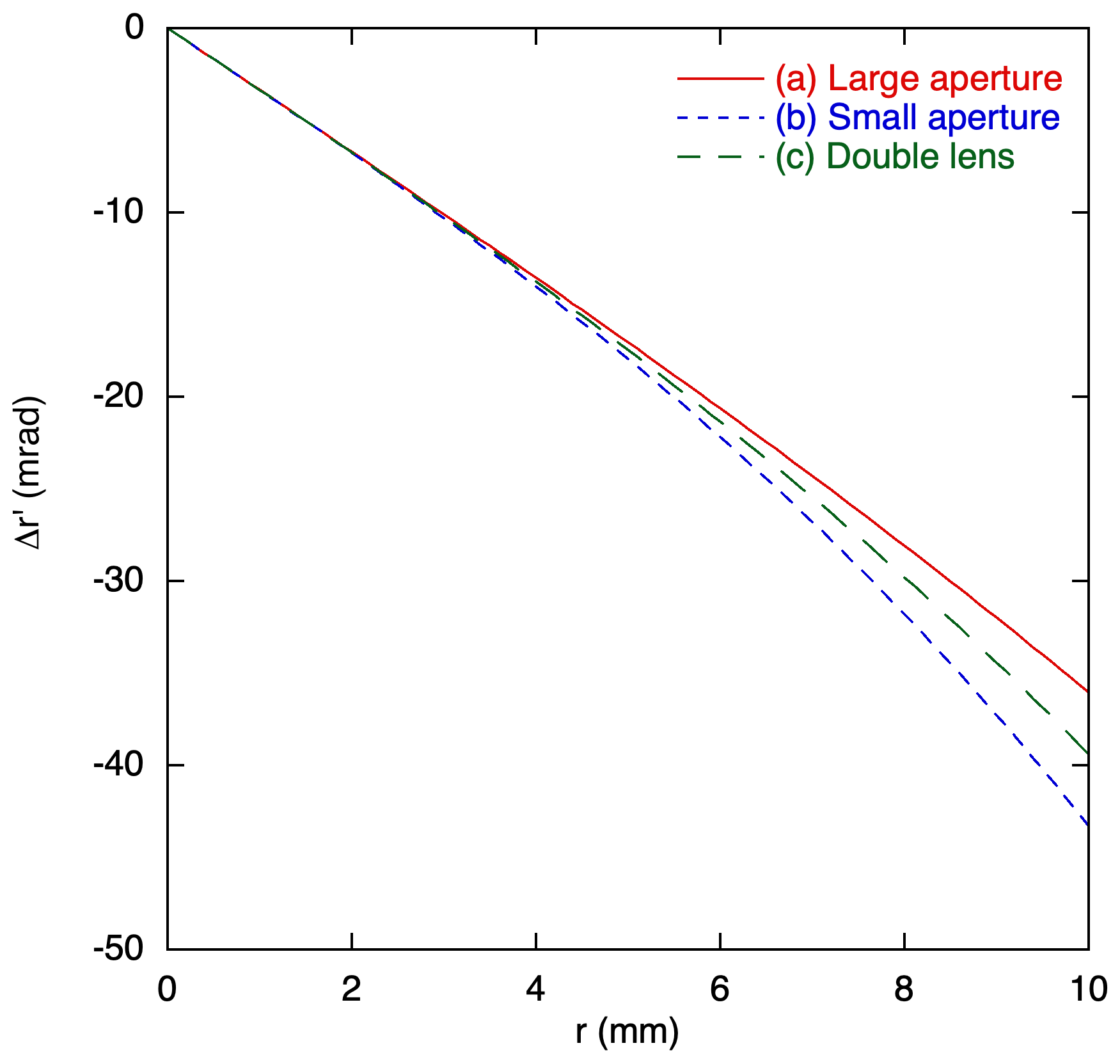}
\caption{\label{fig:phasespace}Phase space profiles of electron beams after passing through the solenoid lenses.}
\end{figure}

The $r$-$\Delta r'$ phase space profiles after passing through the solenoid lens can be calculated using Eq.(\ref{eqn:drd3}). The integral of the square of the magnetic field and derivative were calculated using field data shown in Fig.~\ref{fig:field}. It is assumed that the beam energy is 500 keV and the radial position of electron in the field is constant. It is clearly shown in Fig.~\ref{fig:phasespace} that the nonlinearity of the profile curve was distinctive for small aperture and double lenses. 

\begin{figure}[ht]
\includegraphics[width=10cm]{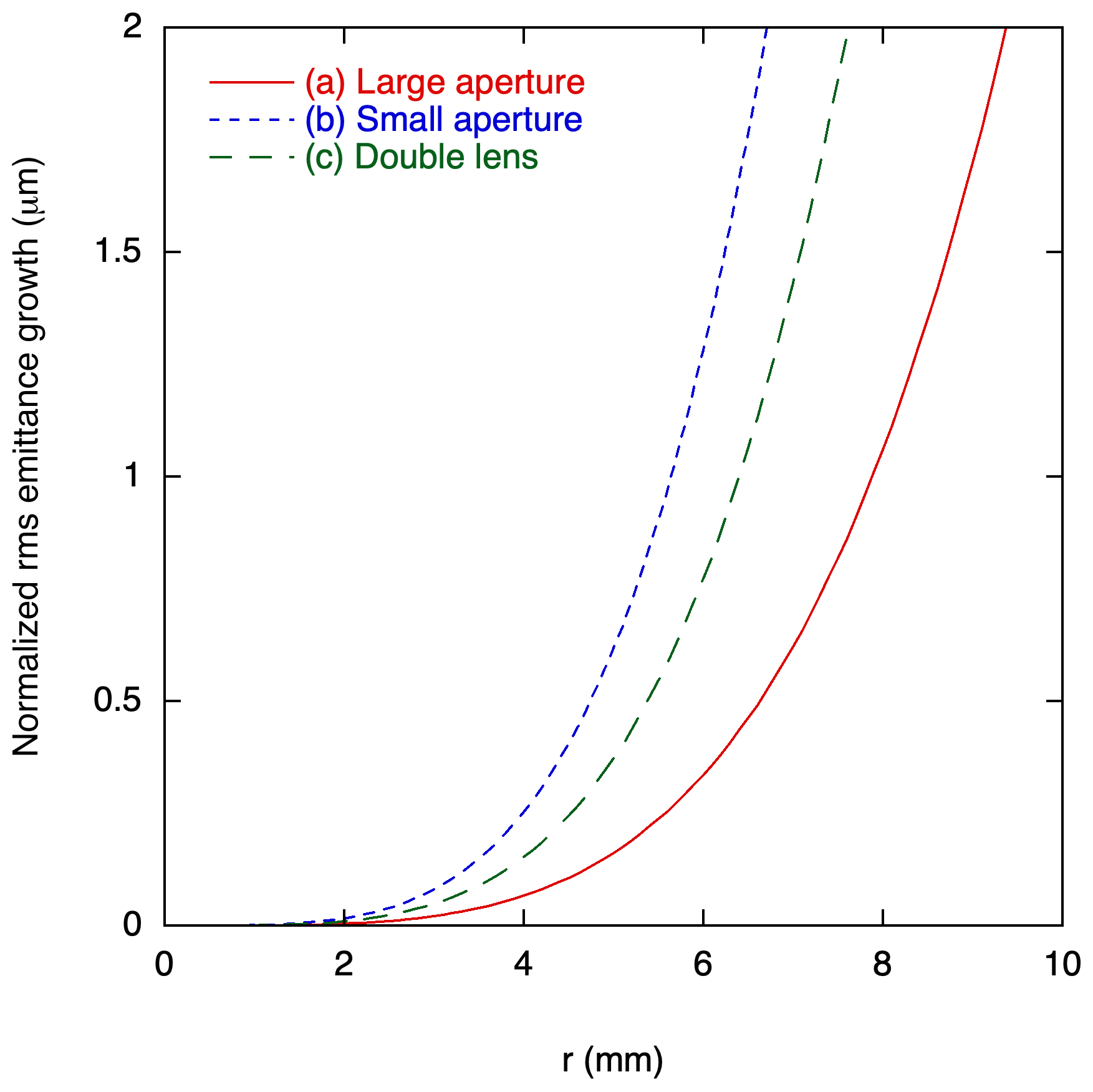}
\caption{\label{fig:emittance}Emittance growths owing to solenoid lens aberration as a function of the beam radius.}
\end{figure}

The emittance growth was calculated using Eq.(\ref{eqn:emittance2}) and is shown in Fig.~\ref{fig:emittance}. It can be seen that the emittance growth is significantly suppressed by enlarging the pole piece aperture. In the SACLA injector, the beam diameter is controlled to less than $\sim\phi$5 mm. In this case, the emittance growth is less than 0.01 $\mu$m per each lens. Therefore, it is predicted that the total emittance growth should not exceed 0.2 $\mu$m. 

The abovementioned emittance growth is caused by geometric distortion of the linearly distributed phase space profile, i.e., a non-thermal phenomenon. A beam extracted from a hot cathode has a thermal emittance $\varepsilon_{n,rms}^{th}$ determined by its temperature and radius. The net emittance is given by their root-sum-square of:
	\begin{eqnarray}
	\varepsilon_{n,rms}^{net}=\sqrt{{\varepsilon_{n,rms}^{th}}^2+{\Delta\varepsilon_{n,rms}^{lens}}^2}. \nonumber
	\end{eqnarray} 
The CeB$_6$ gun generates a beam with a low thermal emittance of 0.6 $\mu$m \cite{PhysRevSTAB.10.020703}. Therefore, the increment of net emittance is $\sim$0.03 $\mu$m. This value is negligibly small for the XFEL injector. It should be noted that $\varepsilon_{n,rms}^{net}$ represents the sliced beam emittance, which is important in achieving FEL gain saturation for XFEL. Increment of the projected emittance owing to time dependent energy dispersion in the bunched beam is not mentioned here. 

According to the preceding discussion, a smaller beam size is favorable in minimizing lens aberration; however, the space charge force in the beam becomes stronger. The equation of motion of an electron in a space charge dominated beam is described by
	\begin{eqnarray}
	\frac{d^2r}{dz^2}=\frac{2I(r)}{I_0(\beta\gamma)^3}\frac{1}{r}, \nonumber
	\end{eqnarray} 
	\begin{eqnarray}
	I(r)=\int_{0}^{r}J(r_1)2{\pi}r_1dr_1, \nonumber
	\end{eqnarray} 
where $I_0$ is Alf\'en current (17 kA) and $J(r)$ is a beam current density at a radial position $r$. If the current density is nonuniform, the nonlinearity of the beam diverging angle $r'$ is reinforced by the strong space charge. In particular, it becomes severe after the sub-harmonic buncher, where the beam is bunched up. Therefore, the beam size must be tuned as a compromise between minimizing the lens aberration and avoiding emittance growth owing to the nonlinear space charge.  

\section{Summary}

In this study, we derived the approximative solution of an emittance growth owing to the solenoid lens aberration. It was found that the emittance develops strongly when the incident beam has a large diameter. In addition, it was evident that the derivative of longitudinal magnetic field affected the emittance. Using derived formulae, the solenoid lens for the SACLA injector was designed to sustain low beam emittance. Therefore, it can be concluded that emittance growth owing to lens aberration is negligibly small for SACLA.

\begin{acknowledgments}

The author would like to thank Dr. T. Shintake of Okinawa Institute of Science and Technology Graduate University, Dr. A. D. Yeremian of SLAC National Laboratory, and V. A. Goryashko of Uppsala University for valuable discussion during this research. 

\end{acknowledgments}

\bibliography{solenoid}

\end{document}